\documentclass[twocolumn]{aastex631}
\usepackage{multirow}

\usepackage{hhline} 

\usepackage{amsmath,amssymb,bm}
\usepackage{soul}
\usepackage{bigints}
\DeclareMathOperator{\real}{Re}

\newcommand{\expval}[1]{\left\langle #1\right\rangle}
\hypersetup{pagebackref, pdfpagemode=FullScreen, colorlinks=true,citecolor=blue,linkcolor=blue}
\begin{document}

\title{Direct Measurements of Synchrotron-Emitting Electrons at Near-Sun Shocks}

\correspondingauthor{I.~C.~Jebaraj}
\email{immanuel.c.jebaraj@gmail.com}

\author[0000-0002-0606-7172]{I.~C.~Jebaraj}
\affil{Department of Physics and Astronomy, University of Turku, 20500 Turku, Finland}

\author[0000-0001-6427-1596]{O. V. Agapitov}
\affil{Space Sciences Laboratory, University of California, Berkeley, CA 94720, USA}
\affil{Astronomy and Space Physics Department, National Taras Shevchenko University of Kyiv, 03127 Kyiv, Ukraine}

\author[0000-0003-1236-4787]{M. Gedalin}
\affiliation{Department of Physics, Ben Gurion University of the Negev, Beer-Sheva 84105, Israel}

\author[0000-0002-2238-109X]{L. Vuorinen}
\affil{Department of Physics and Astronomy, University of Turku, 20500 Turku, Finland}
\affil{Department of Physics and Astronomy, Queen Mary University of London, London E1 4NS, United Kingdom}

\author[0000-0003-0876-8391]{M. Miceli}
\affil{Dipartimento di Fisica e Chimica E. Segrè, Università degli Studi di Palermo, Palermo, 90134, Italy}

\author[0000-0002-3298-2067]{R. Vainio}
\affil{Department of Physics and Astronomy, University of Turku, 20500 Turku, Finland}

\author[0000-0002-0978-8127]{C. M. S. Cohen}
\affil{California Institute of Technology, Pasadena, CA 91125, USA}

\author[0000-0001-8307-781X]{A. Voshchepynets}
\affil{Department of System Analysis and Optimization Theory, Uzhhorod National University, 88000 Uzhhorod, Ukraine}
\affil{Space Sciences Laboratory, University of California, Berkeley, CA 94720, USA}

\author[0000-0001-6589-4509]{A. Kouloumvakos}
\affil{The Johns Hopkins University Applied Physics Laboratory, Laurel, MD 20723, USA}

\author[0000-0003-3903-4649]{N. Dresing}
\affil{Department of Physics and Astronomy, University of Turku, 20500 Turku, Finland}

\author[0000-0003-1175-7124]{A. Marmyleva}
\affil{University of Helsinki, 00014 Helsinki, Finland}

\author[0000-0002-6809-6219]{V. Krasnoselskikh}
\affiliation{LPC2E/CNRS, UMR 7328, 45071 Orléans, France}
\affil{Space Sciences Laboratory, University of California, Berkeley, CA 94720, USA}

\author[0000-0002-8110-5626]{M. Balikhin}
\affil{University of Sheffield, Sheffield S10 2TN, United Kingdom}

\author[0000-0003-4501-5452]{J. G. Mitchell}
\affil{Heliophysics Science Division, NASA Goddard Space Flight Center, Greenbelt, MD 20771, USA}

\author[0000-0001-9178-5349]{A. W. Labrador}
\affil{Heliophysics Science Division, NASA Goddard Space Flight Center, Greenbelt, MD 20771, USA}

\author[0000-0001-6344-6956]{N. Wijsen}
\affil{Centre for mathematical Plasma Astrophysics, KU Leuven, 3001 Leuven, Belgium}

\author[0000-0001-6590-3479]{E. Palmerio}
\affil{Predictive Science Inc., San Diego, CA 92121, USA}

\author[0000-0001-6016-7548]{L. Colomban}
\affil{Space Sciences Laboratory, University of California, Berkeley, CA 94720, USA}

\author[0000-0003-1175-7124]{J. Pomoell}
\affil{University of Helsinki, 00014 Helsinki, Finland}

\author[0000-0002-4489-8073]{E.~K.~J. Kilpua}
\affil{University of Helsinki, 00014 Helsinki, Finland}

\author[0000-0002-1573-7457]{M. Pulupa}
\affil{Space Sciences Laboratory, University of California, Berkeley, CA 94720, USA}

\author[0000-0002-2011-8140]{F. S. Mozer}
\affil{Space Sciences Laboratory, University of California, Berkeley, CA 94720, USA}

\author[0000-0003-2409-3742]{N. E. Raouafi}
\affil{The Johns Hopkins University Applied Physics Laboratory, Laurel, MD 20723, USA}

\author[0000-0001-6160-1158]{D. J. McComas}
\affil{Department of Astrophysical Sciences, Princeton University, Princeton, NJ 08544, USA}

\author[0000-0002-1989-3596]{S. D.~Bale}
\affil{Physics Department, University of California, Berkeley, CA 94720, USA}
\affil{Space Sciences Laboratory, University of California, Berkeley, CA 94720, USA}

\begin{abstract}

In this study, we present the first-ever direct measurements of synchrotron-emitting heliospheric traveling shocks, intercepted by the Parker Solar Probe (PSP) during its close encounters. Given that much of our understanding of powerful astrophysical shocks is derived from synchrotron radiation, these observations by PSP provide an unprecedented opportunity to explore how shocks accelerate relativistic electrons and the conditions under which they emit radiation. The probe's unparalleled capabilities to measure both electromagnetic fields and energetic particles with high precision in the near-Sun environment has allowed us to directly correlate the distribution of relativistic electrons with the resulting photon emissions. Our findings reveal that strong quasi-parallel shocks emit radiation at significantly higher intensities than quasi-perpendicular shocks due to the efficient acceleration of ultra-relativistic electrons. These experimental results are consistent with theory and recent observations of supernova remnant shocks and advance our understanding of shock physics across diverse space environments.

\end{abstract}


\section{Introduction} \label{sec:intro}

Collisionless shocks are among the most ubiquitous and strongly nonlinear systems in plasma, spanning spatial scales from laboratory settings to galaxy clusters \citep{Sagdeev66, Galeev76, Kennel85, Lembege04, Krasnoselskikh13, Agapitov23, Miceli23}. In the shock reference frame, most part of the directed flow’s kinetic energy is converted into plasma heating, particle acceleration, magnetic compression, and turbulence. Understanding the mechanisms behind this energy conversion presents a significant challenge. Electron heating and acceleration to relativistic energies are key channels of energy redistribution, and the analysis of the electromagnetic (EM) radiation from these electrons is crucial for remote study of astrophysical shocks \citep{Vink20book}. 

Of particular interest is synchrotron radiation emitted by relativistic electrons accelerated at shock waves \citep{Schwinger49, Born64, Ginzburg64}, which is the most common EM emission observed from supernova remnant (SNR) shocks \citep{Bykov04, Helder09}, spanning from radio to X-ray wavelengths. The energy of these electrons is given by \(E_\mathrm{e} = \gamma m_\mathrm{e} c^2\), where \(\gamma\) (\({\gg}1\)) is the Lorentz factor, \(m_\mathrm{e}\) the electron mass, and \(c\) the speed of light. Consequently, synchrotron emission is highly beamed within an angle \(\theta \sim 1/\gamma\) around the electron's velocity direction. The emission power peaks at the critical frequency, \(\omega_\mathrm{m} \sim \Omega_\mathrm{e} \gamma^2\), where \(\Omega_\mathrm{e} = eB/m_\mathrm{e} c\) is the non-relativistic electron gyrofrequency, with \(e\) being the electron charge and \(B\) the ambient magnetic field \citep{Ginzburg1979Theoretical-Phy, Longair92}. The synchrotron spectrum is broadband, with a width \(\Delta \omega \sim \omega_\mathrm{m}\), and the total emitted power scales as \(\mathcal{P} \sim \gamma^2\). However, this emission typically arises from a distribution of electrons, often following a power law with an exponent \(\delta\).

The acceleration of such power-law distributed electrons, which emit broadband radiation, is often described by diffusive shock acceleration (DSA) theory \citep{Krymskii77, Axford77, Bell78, Blandford78}. In the strong shock limit (gas compression ratio, \(r_\mathrm{gas} = 4\)), DSA predicts a power-law electron energy distribution with \(\delta = (r_\mathrm{gas} + 2)/(r_\mathrm{gas} - 1) = 2\) \citep{Malkov01, Bykov19book}. A typical synchrotron spectrum produced by such an electron distribution has two main features: a low-frequency turnover (\(\omega_\mathrm{c}\)) separating optically thick and thin regimes \citep{Ginzburg64, Rybicki79}, and a high-frequency exponential cutoff due to the maximum electron energy, \(\omega_\mathrm{max} \sim \Omega_\mathrm{e} \gamma_\mathrm{max}^2\), where \(\gamma_\mathrm{max}\) is the highest Lorentz factor in the distribution.

In the optically thin regime, the synchrotron spectrum follows a power law with an exponent \(\alpha \sim 0.5\), which directly relates to the electron energy distribution (\(\delta = 2\)) via \(\alpha = (\delta - 1)/2\). This relationship is traditionally used to estimate particle distributions in astrophysical sources such as SNRs, though variations are common \citep{Green19}. Despite these insights, uncertainties remain about the detailed distribution of relativistic electrons, their acceleration mechanisms, and the specific conditions under which they emit radiation. The universality of DSA theory, particularly its scale invariance in strong shocks (\(r_\mathrm{gas} = 4\)), suggests that understanding synchrotron radiation in interplanetary (IP) shocks could offer a bridge to understanding SNR shocks. Such a bridge between IP and SNR shocks has long been sought for, with the heliosphere often considered a practical laboratory for studying remote astrophysical objects \citep{Kennel85,Terasawa03}. Observations of SNR shocks offer unique insights into the analysis of global shock structure through high-resolution radio astronomy, which allows for probing different regions of a shock with GeV electrons along various lines of sight. However, certain factors must be considered, such as the vast difference in system size, with SNR shocks being more than 6 orders of magnitude larger and their lifetime is much longer. Similarly, the energy they carry is at least 15 orders of magnitude greater than that of the most powerful IP shocks \citep{Vink20book}. Furthermore, the nonstationarity of acceleration, e.g. in traveling IP shocks,  may substantially affect the particle spectrum, making comparison with SNR shocks more difficult.

A combination of remote-sensing observations of synchrotron emission and \textit{in situ} measurements of the emitting electrons can address key questions about the conditions that produce accelerated electrons capable of synchrotron radiation. This approach could help reconcile differences between strong IP and SNR shocks, where system size and the upstream ion bulk energy in the shock frame are the primary distinctions. Earlier studies suggested that heliospheric synchrotron emission is generated by energetic electrons trapped within the coronal mass ejections driving the shock, rather than by shock-accelerated electrons \citep{Bastian07,Pohjolainen13}. However, more recent work has shown that traveling IP shocks can accelerate electrons to relativistic energies \citep{Nasrin23, Jebaraj23, Jebaraj24}.

In this study, we present the first-ever direct observations of synchrotron-emitting relativistic (\(\gamma \sim 2\)--\(6\) or \(E_\mathrm{e} \sim 1\)--\(3\)~MeV) and ultra-relativistic (\(\gamma > 10\) or \(E_\mathrm{e} > 5\) MeV) electrons produced by fast IP shocks observed by NASA’s Parker Solar Probe \citep[PSP;][]{Fox2016}. We utilize continuous, high-fidelity electric field measurements in the radio wavelengths from the Radio Frequency Spectrometer \citep[RFS;][]{Pulupa2017}, part of the FIELDS instrument suite \citep{Bale16}. These observations are compared with the electron distributions measured at the time of shock arrival by the Low and High Energetic Particle Instruments \citep[EPI-Lo \& EPI-Hi, respectively;][]{Hill17, Wiedenbeck17}, part of the Integrated Science Investigation of the Sun (IS\(\odot\)IS) suite \citep{McComas16}. The first shock (S1) was quasi-perpendicular and detected during PSP’s 13th close approach to the Sun (Encounter 13) on September 5, 2022, at 15.1 \(R_{\odot}\) \citep{Romeo23}. The second shock (S2) was near-parallel and detected 49 \(R_{\odot}\) away during Encounter 15 on March 13, 2023 \citep{Jebaraj24}. In the following sections, we present our observations of synchrotron emission, analyze the spectral characteristics and polarization, and examine the distribution of the emitting electrons.

\section{Results from the analysis}

\begin{figure*}[ht!]
\centering
\includegraphics[width=1\textwidth]{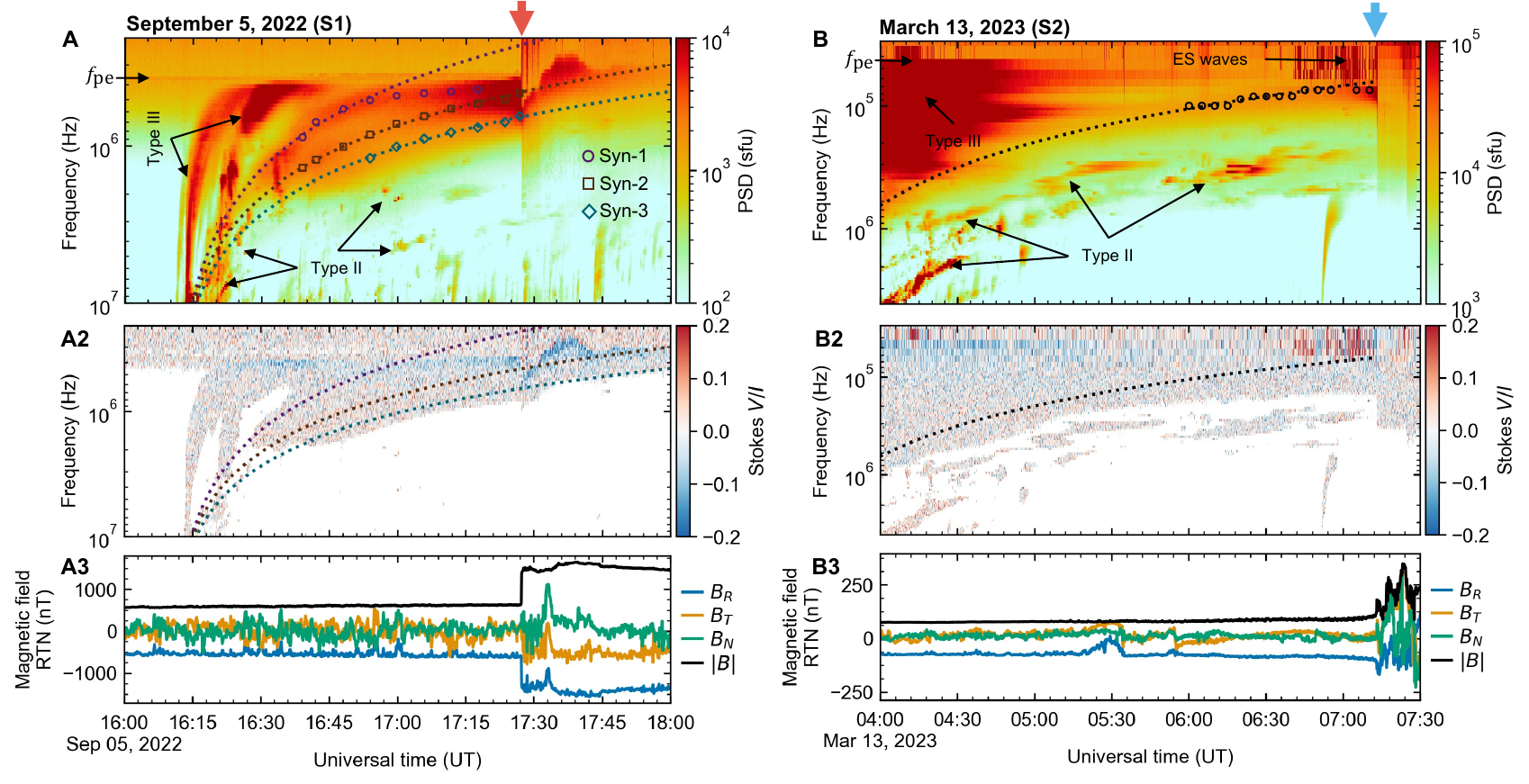}
\caption{Panels A and B (top row) show the dynamic spectra of electric field power, \(E(f,t)^2\), in the radio frequency range, measured before and at the shock crossing on September 5, 2022 (S1) and March 13, 2023 (S2), respectively. The units are in solar flux units (1 sfu = \(10^{-22}\) W m\(^{-2}\) Hz\(^{-1}\)). The arrival of both shocks at PSP is marked by the red (S1) and blue (S2) arrows on top of the figure. The synchrotron emissions, the coherent plasma emissions (type II and type III), and the local \(f_\mathrm{pe}\) are marked. Panels A2 and B2 (middle row) show the degree of circular polarization normalized by total intensity (Stokes-\(V/I\)) (see text for details). Panels A3 and B3 (bottom row) show the magnetic field components and magnitude.}\label{fig1}
\end{figure*}

\subsection{Diffuse and self-absorbed continuum-like emission} \label{sec:res1}

Figures \ref{fig1}A and \ref{fig1}B display the dynamic radio spectra of electric field intensity from transverse EM waves, \(E(f,t)^2\), where \(f = \omega/2\pi\) (in Hz), recorded by the high (HFR; 19.2--1.6 MHz) and low (LFR; 1.6--0.001 MHz) frequency receivers of FIELDS/RFS onboard PSP, both before and during the shock crossings on September 5, 2022 (S1), and March 13, 2023 (S2), respectively. The units of measurement are solar flux units (1 sfu = \(10^{-22}\) W m\(^{-2}\) Hz\(^{-1}\)), normalized to 1~au, assuming that the intensity decreases with distance (\(r\)) following \(r^{-2}\). These EM waves, emitted by electrons at some distance from the spacecraft, propagate toward the observer at the speed of light, appearing as distinct maxima in the electric field intensity. The frequencies of these maxima change over time, and the rate of change, hereafter referred to as ``drift", reflects variations in plasma density and magnetic field gradients in the emission region. In regions with negative gradients in plasma density and magnetic field, both \(\omega_\mathrm{pe}\) (\(= \sqrt{(n_\mathrm{e} e^2)/(4\pi \epsilon_0 m_\mathrm{e})}\), where \(n_\mathrm{e}\) is electron density and \(\epsilon_0\) is the permittivity of free space) and \(\Omega_\mathrm{e}\) will show negative gradients. As a result, analyzing spectral drift rates and assuming a model for the emission mechanism, properties of the emitting electrons and the surrounding medium can be inferred. The electron plasma frequency at the observer's location, \(f_\mathrm{pe} = \omega_\mathrm{pe}/2\pi\), is annotated in both Figures~\ref{fig1}A and \ref{fig1}B. Below \(f_\mathrm{pe}\), EM waves become evanescent, resulting in a sharp emission cutoff regardless of where the emission originates.

Figures \ref{fig1}A and \ref{fig1}B also annotate common coherent emissions generated by plasma instabilities \citep{kaplan1969plasma,Papadopoulos78,Krasnoselskikh79,Vosh15}. This includes type II bursts, produced by subrelativistic electrons accelerated at shocks. This emission is characterized by a narrow relative frequency bandwidth \citep[\(\Delta f/f = (f_\mathrm{high} - f_\mathrm{low})/(f_\mathrm{center}) < 0.3\), where \(f_\mathrm{center} = (f_\mathrm{high} + f_\mathrm{low})/2\);][]{Cairns03, Jebaraj20}. In the spectra, type II bursts appear as chains of intense, fragmented emissions with frequency drift corresponding to the speed of the shock region where they originate. Additionally, type III bursts are generated by subrelativistic electron beams propagating along open magnetic field lines, with \(\Delta f/f \lesssim 0.5\) \citep{Suzuki85book, Jebaraj23l}. Type III bursts, the most intense heliospheric radio emissions, exhibit rapid spectral drift rates, reflecting the high speed of the emitting electrons. While these emissions and their relationship to solar transient shocks have been extensively studied over the past five decades, this study focuses on spectral features beyond these well-known emissions.

\textit{September 5, 2022 (S1)}: In Figure \ref{fig1}A, we identify three distinct emissions that deviate from the characteristics of the aforementioned coherent emissions, annotated as Syn-1, Syn-2, and Syn-3. 
Syn-1 (circle markers) starts at \(\sim\)5 MHz at 16:20 UT and drifts down to the local \(f_\mathrm{pe} \sim 400\) kHz by 17:00 UT, with some parts observed until shock arrival at 17:27 UT.
Syn-2 (square markers) begins at \(\sim 2\) MHz at 16:30 UT and drifts down to \(f_\mathrm{pe}\) during shock arrival at 17:27 UT.
Syn-3 (diamond markers) starts at \(\sim 2\) MHz at 16:50 UT and similarly drifts down to \(f_\mathrm{pe}\) by 17:27 UT.
These emissions are characterized as diffuse, broadband emissions with \(\Delta f/f \sim 1\), and the associated maxima (annotated with distinct markers) correspond to the turnover frequency (\(f_\mathrm{c} = \omega_\mathrm{c}/2\pi\)). The features are indicative of synchrotron emission with self-absorption \citep[][]{Nindos20}. It is worth noting that some emission is seen continuing in the post-shock region up till 17:45 UT. Much of it is evanescent due to the high \(f_\mathrm{pe}\) in the post-shock plasma. Analysis in Appendix \ref{app:TDS} shows that the measured electric field intensity corresponds to EM waves with phase speeds close to \(c\), with no detection of Langmuir waves. 

\textit{March 13, 2023 (S2)}: In Figure \ref{fig1}B, an emission starts at 05:20 UT from \(\sim 200\) kHz and extends until shock arrival at 07:13 UT. This emission is also characterized as diffuse and broadband with \(\Delta f/f \sim 1\), with \(f_\mathrm{c}\) annotated with circle markers. A similar analysis of the EM waves near the shock arrival at 07:12 UT is presented in Appendix \ref{app:TDS}. The estimated phase speeds for the waves above \(f_\mathrm{pe}\) were close to \(c\). However, unlike S1, several electrostatic waves were detected at a small fraction of \(f_\mathrm{pe}\), annotated as ``ES waves" in Figure \ref{fig1}B. These waves are not expected to contribute to the generation of coherent plasma emissions \citep{Lobzin05}.

\subsection{Emission is depolarized in inhomogeneous plasma} \label{sec:res2}

Synchrotron radiation from a homogeneous medium is highly polarized \citep[][]{Ginzburg64}. EM radiation is emitted as a combination of two transverse wave modes, distinguished by the orientation of the wave electric field (\(\mathbf{E}\)) relative to the background magnetic field (\(\mathbf{B}\)). These modes are the ordinary (\(o\)-mode, \(\mathbf{E} \parallel \mathbf{B}\)) and extraordinary (\(x\)-mode, \(\mathbf{E} \perp \mathbf{B}\)) waves. In a cold, unmagnetized plasma, the propagation of \(o\)- and \(x\)-modes is independent of each other, but this is not the case in a magnetoactive plasma. The transfer of these wave modes through a plasma is best described using the Stokes parameters: \(I\), \(Q\), \(U\), and \(V\) \citep[][]{chandrasekhar1947transfer,Ramaty69}. A short description of the Stokes parameters and the means of estimating them from data can be found in Appendix~\ref{app:stokes}. When wave propagation is quasi-parallel to \(\mathbf{B}\), the Stokes parameters simplify such that \(Q = U \rightarrow 0\). As a result, \(o\)- and \(x\)-mode waves are circularly polarized in opposite directions, and Stokes-\(V\) is given by the difference between the intensities of the \(o\)- and \(x\)-modes (\(V = I_\mathrm{O} - I_\mathrm{X}\)), resulting in circular polarization in the direction of the dominant mode. In the case of quasi-perpendicular wave propagation relative to \(\mathbf{B}\), Stokes-\(Q = I_\mathrm{O} - I_\mathrm{X}\), while \(U = V \rightarrow 0\), leading to predominantly linear polarization. 

The polarization conventions described above apply strictly to the region where emission originates. If \(\mathbf{B}\) changes between the emitter and observer, the polarization can be altered or lost. Stokes-\(V\) may be converted to Stokes-\(Q\) (and vice versa) due to Faraday conversion, potentially reducing or reversing the observed circular polarization \citep[][]{Sitenko66,Zhelez68}. Linear polarization can undergo Faraday rotation, where the polarization plane rotates during propagation, possibly leading to depolarization if the magnetic field is complex \citep[][]{Sitenko66,Ginzburg69,Melrose71}. Precise measurements of polarization is impossible, as the waves may become depolarized while propagating through a plasma with random fluctuations. Consequently, for a remote observer, it may differ significantly from what it was at the region where it is emitted.

Synchrotron radiation is typically emitted over a range of angles with respect to \(\mathbf{B}\), making the \(x\)-mode dominant for the observer. This also makes the emission linearly polarized in the \(x\)-mode when observed along the direction of the electron's velocity or from any oblique angles to \(\mathbf{B}\). Thus, the degree of circular polarization is secondary and increases with \(\theta \sim 1/\gamma\), or when viewed from oblique angles within \(\theta\). This can be reformulated in terms of Stokes parameters as \(V/Q \sim 1/\gamma\). The handedness of circular polarization for the \(o\)-mode is strictly clockwise (right-handed) when the magnetic field is pointing inwards (\(B_\mathrm{r}<0\)), and counterclockwise (left-handed) when it is pointing outwards (\(B_\mathrm{r}>0\)). For the \(x\)-mode, the handedness is reversed.

The emissions from both S1 and S2 have negligible degree of linear polarization and are not presented here. The degree of circular polarization  \(\left((I_\mathrm{O} - I_\mathrm{X})/(I_\mathrm{O} + I_\mathrm{X}) = V/I \right)\) for both events is depicted in Figures \ref{fig1}A2 and \ref{fig1}B2. 

\textit{September 5, 2022 (S1):} We observed clear circular polarization for Syn-2 and Syn-3 when the emission was close to the local \(f_\mathrm{pe}\) during the shock arrival between 17:15 and 17:27 UT. Given that \(V/Q \approx 1/\gamma\), we used the degree of circular polarization (\(V/I = 15\)--\(25\%\)) to estimate that the emission originated from electrons with \(\gamma \sim 4\)--\(7\). Figure \ref{fig1}A3 shows the magnetic field components and magnitude, revealing that S1 is polarized in the left-hand sense while \(B_\mathrm{r} < 0\) (inward-pointing), consistent with emission in the \(x\)-mode as predicted by theory. While remote observations can carry significant uncertainties when correlating with magnetic field data, our measurements were made with the emitting region in close proximity to the observer, thereby reducing much of this uncertainty.

\textit{March 13, 2023 (S2):} The observations presented in Figure \ref{fig1}B2 show no signs of circular polarization. This lack of polarization is likely due to a combination of observer position and depolarization caused by a randomly inhomogeneous medium. If the conditions at and around the emitting region are turbulent at scales corresponding to the wavelength of the EM waves, they may be depolarized within the emitting region. Even in the presence of large scale inhomogeneities, Faraday effect may significantly depolarize the EM waves. Indeed, it was demonstrated by \cite{Jebaraj24} that inhomogenities across a wide range of scales were present. 

\begin{figure*}[htbp]
\centering
\includegraphics[width=0.99\textwidth]{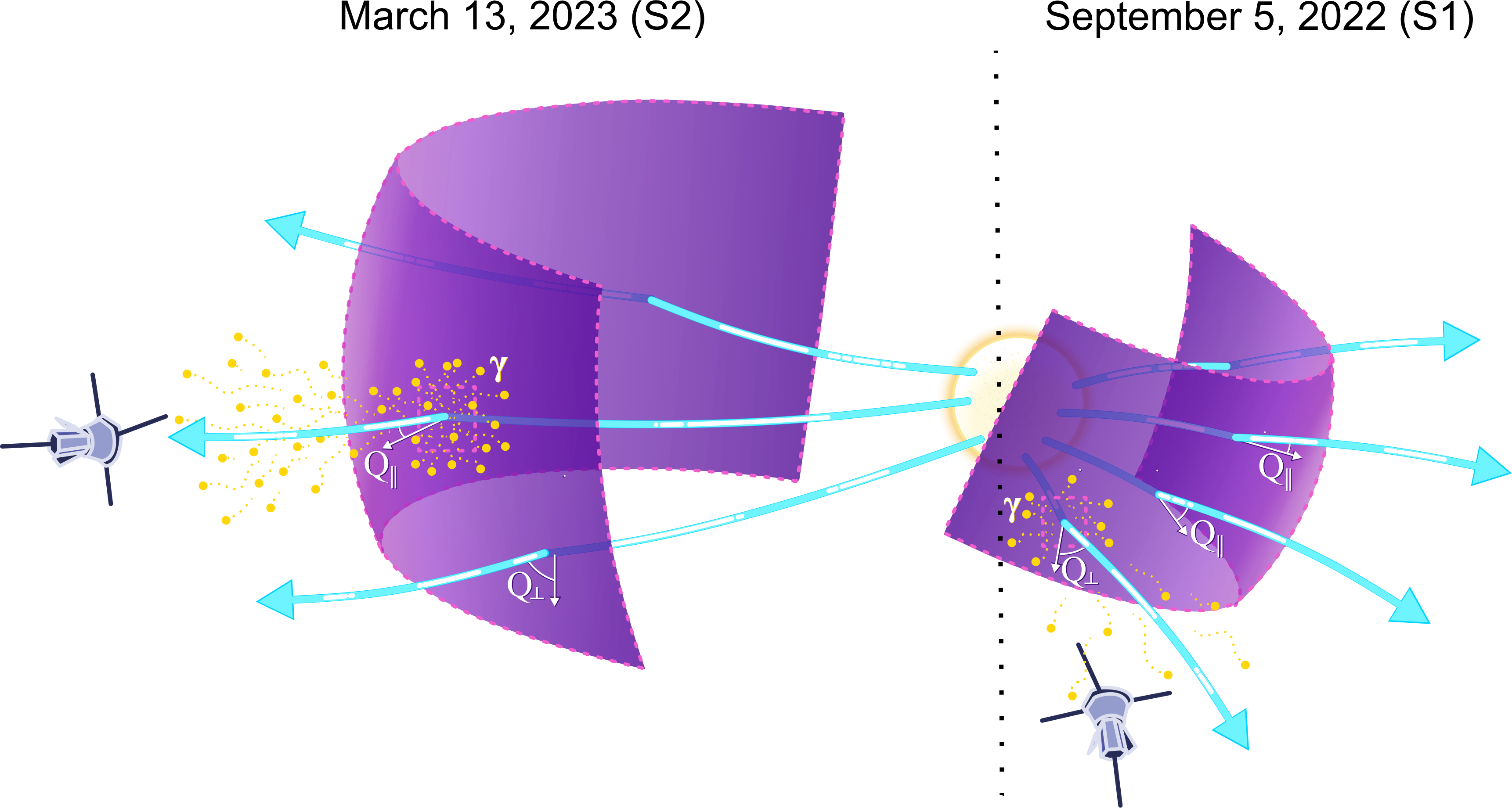}
\caption{Schematic of the synchrotron emission from S1 and S2. The curved purple front represents the traveling shock wave, with magnetic field lines shown in blue. The shock geometry along the reference field lines is labeled as $Q_\perp$ (quasi-perpendicular) and $Q_\parallel$ (quasi-parallel). The finite region on the shock surface where relativistic electrons are accelerated and subsequently emit synchrotron radiation is indicated by the red square. The emitted photons are represented by yellow markers and the notation $\gamma$.}
\label{fig_sch}
\end{figure*}

\subsection{Intrinsic brightness increases as shock approaches PSP}

The characteristics of the synchrotron emission may also be assessed taking into account the radiation transfer from the emitting electrons towards the observer \citep[][]{chandrasekhar1947transfer}. Under the Rayleigh--Jeans approximation (\(hf \ll k_\mathrm{B} T_\mathrm{B}\), where \(h\) is Planck's constant, \(f\) is the frequency of radiation in Hz, and \(k_\mathrm{B}\) is the Boltzmann constant), \(T_\mathrm{B} \approx (c^2 I_f)/(2 k_\mathrm{B} f^2)\), where \(I_f\) is the emission intensity. The intrinsic brightness of the source or brightness temperature \(T_\mathrm{B}\) of synchrotron emissions cannot exceed the limit set by inverse Compton scattering \citep[\(10^{12}\)~K at 1 GHz,][]{Kellerman69,Rybicki79}. Quantifying the solid angle subtended by the source (\(\Omega\) in units of steradians) allows us to estimate \(T_\mathrm{B}\) (see  \eqref{eq:brithnesstempsolid} in Appendix \ref{app:Brightnesstemp}).  

\textit{September 5, 2022 (S1):} In Figure \ref{fig1}A, the maxima of Syn-1, Syn-2, and Syn-3 are annotated by the circle, square, and diamond markers, respectively. They correspond to the self-absorption cutoff or the \(f_\mathrm{c}\). If the source of the synchrotron emission are the relativistic electrons accelerated at the shock front that rapidly approaches the probe, then \(\Omega\) increases at a rate determined by the shock speed. This suggests that the presence of different synchrotron emissions (Syn-1, Syn-2, and Syn-3) are likely due to different electron populations at different regions of the shock, which propagate at different speeds. Assessing Syn-2 and Syn-3, we find that their intensity increase with decreasing frequency indicating a rapid increase in \(\Omega\), such that \(\Omega \rightarrow 2\pi\).

The effect of \(\Omega\) is such that as the region of the shock accelerating the electron gets closer and closer, emission from more and more electrons is received. Such an effect would also be seen in \(T_\mathrm{B}\) which was estimated for Syn-1, Syn-2, and Syn-3 in Appendix~\ref{app:Brightnesstemp}. The result suggests that none of them exceed the inverse Compton scattering limit of \(1.3 \times 10^{13}\)~K. However, the \(T_\mathrm{B}\) of Syn-2 and Syn-3 increases rapidly just before the shock arrival at 17:27 UT indicating that \(\Omega \rightarrow 2\pi\), i.e. emission originates from regions in close proximity to the shock front traversed by the spacecraft. A schematic of the likely shock encounter is shown on the right in Fig. \ref{fig_sch}, where the probe encounters the shock flanks which is quasi-perpendicular. The probe is positioned in close proximity to the emitting region (marked by the red square on the shock surface), allowing it to receive portions of the synchrotron radiation.

\textit{March 13, 2023 (S2):} The intensity of the emission increases as the shock approaches. Quantifying this using the solid angle \(\Omega\) we find that \(T_\mathrm{B}\) remains relatively stable until 06:55 UT, after which it increases exponentially as the shock nears the spacecraft, corresponding to \(\Omega \rightarrow 2\pi\).  Notably, \(T_\mathrm{B}\) exceeds the inverse Compton limit of \(1.7 \times 10^{13}\) K, suggesting that \(\Omega = 2\pi\) at 07:10 UT. This indicates that the spacecraft was inside the region where electron acceleration and subsequent emission were being generated. In this scenario, the observer receives emission from a source that covers the full \(4\pi\) steradians (i.e., a fully isotropic source), causing the \(2\pi\) steradian assumption to break down. The schematic on the left in Fig.~\ref{fig_sch} illustrates the likely encounter between the probe and the shock wave. The probe encounters the center of the quasi-parallel shock region, where it receives the bulk of the emitted synchrotron photons, leading to a higher observed emission intensity. As noted in Fig.~\ref{figA2}, the intrinsic brightness of the radiation from S2 is an order of magnitude greater than that of S1, supporting the representation shown in the schematic.

\subsection{Relativistic electron distribution and the optically thin synchrotron spectra} \label{sec:res3}

\begin{figure*}[ht!]
\centering
\includegraphics[width=0.65\textwidth]{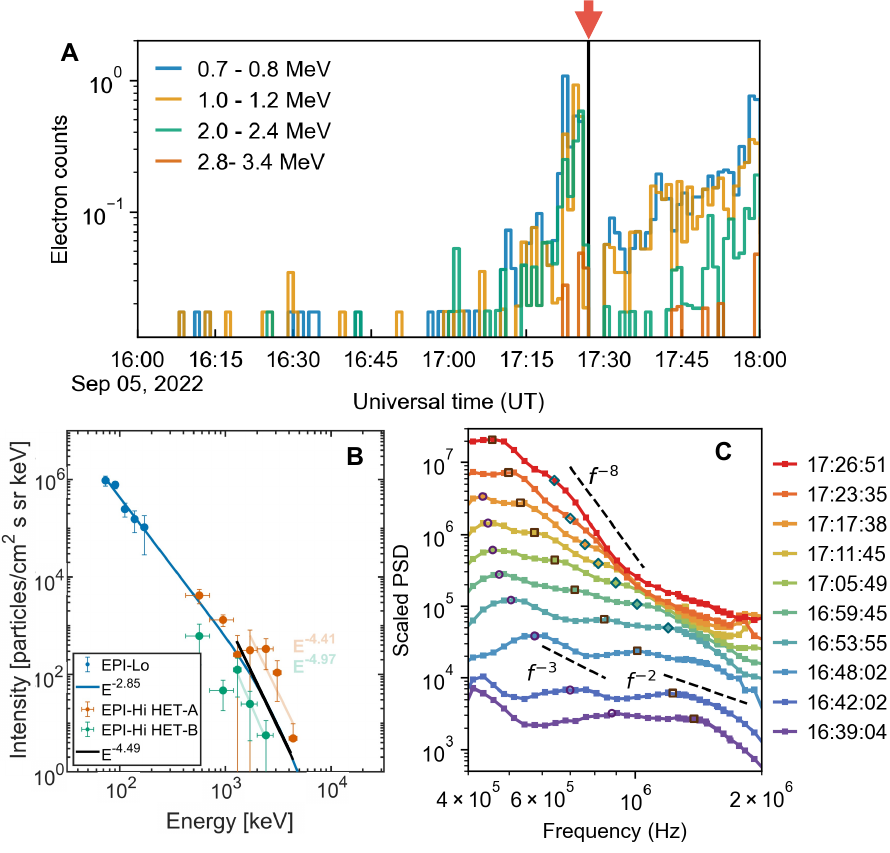}
\caption{Characteristics of the synchrotron emission from S1 and the corresponding electron distribution. Panel A presents the time evolution of the electron intensity in units of counts s\(^{-1}\) observed between 16:00~UT and 18:00~UT. The arrival of the shock at PSP is marked by the vertical black line at 17:27~UT and the red arrow on top. Panel B shows the 10-minute averaged electron energy distribution between 17:17-17:27~UT in units of particles cm\(^{-2}\) s\(^{-1}\) sr\(^{-1}\) keV\(^{-1}\). The different detectors of IS\(\odot\)IS used to construct the electron spectrum are annotated with colored markers. The approximate power law fits and the exponents are also shown in the legend. Panel C shows the intensity-frequency profiles at specific times leading up to the shock arrival, along with approximate power law fits to the optically thin regime. The \(f_\mathrm{c}\) for Syn-1, Syn-2, and Syn-3 are indicated by the distinct markers introduced in Figure \ref{fig1}A. }\label{fig2}
\end{figure*}

The observations of synchrotron emission from shocks that traversed PSP provide a unique opportunity to directly probe the energetic electron distribution and compare it with the synchrotron spectrum. This comparison may be approached from DSA theory as the power law exponents of the synchrotron spectrum (\(\alpha\)) and that of the electrons (\(\delta\)) are directly related through \(\alpha \sim (\delta - 1)/2\). While the shock structure is relatively unimportant for high energy ions undergoing DSA, it plays an important role for electron acceleration \citep{Riquelme2011}. Furthermore, investigating the characteristics of the accelerated electrons such as their anisotropy may also improve our understanding of the process. In the presence of a well-developed ion foreshock, electrons scatter and their pitch angles become diffused, leading to an isotropic distribution \citep[][]{Malkov01}. 

If these shock characteristics play a significant role, we would expect to see differences between S1 and S2, given their distinct properties that have been discussed in \cite{Romeo23} and \cite{Jebaraj24}. S1 was a fast shock that was propagating quasi-perpendicularly to the background \(\mathbf{B}\) and had moderate gas compression, \(r_\mathrm{gas} \sim 2.5\). S2 was also a fast shock, but propagating nearly parallel to \(\mathbf{B}\) with maximum compression, \(r_\mathrm{gas} = 4\).  While shock geometry is not a direct factor in DSA estimations, it plays a crucial role in creating the conditions that enable efficient particle acceleration via DSA, such as the development of a wave foreshock \citep[][]{Drury83,Vainio03}. The formation of a foreshock is particularly effective in quasi-parallel shocks due to ion-streaming instabilities, which generate waves that act as scattering centers for both ions and relativistic electrons, thereby enhancing DSA \citep{Vainio99a}. According to DSA theory, the predicted electron energy spectral index \(\delta\) would be around 3 for S1 and closer to 2 for S2, corresponding to synchrotron spectral slopes of \(\alpha \sim 1\) for S1 and \(\alpha = 0.5\) for S2. However, a limitation of DSA's applicability to electrons is that only those with \(\gamma \gtrsim 2\) or \(E_\mathrm{e} \gtrsim 1\) MeV can participate effectively, as these electrons are able to interact with the waves generated by ions \citep[][]{Zank92,Malkov01}. The technical requirement for electron involvement in DSA is less stringent than for their injection into it. The primary condition is that their speed must exceed that of the shock along the field line, which imposes additional constraints on DSA efficiency in quasi-perpendicular shocks, even in the presence of ion-scale waves.

While this study does not specifically address how and why sub-relativistic electrons reach relativistic energies, the process remains poorly understood and presents a barrier to fully understanding electron acceleration. Previous studies have shown that electron energization is linked to the scale of the electromagnetic fields at the shock transition layer \citep{Balikhin93,Balikhin98}, which determines the contributions from both adiabatic and non-adiabatic processes \citep{Gedalin20}. In the case of S1, where the shock is quasi-perpendicular, much of the sub-relativistic electron acceleration is likely to be adiabatic \citep[][]{Balikhin89,Jebaraj23l}, although non-adiabatic processes may also play a role when the shock is super-critical \citep{gedalin1995demagnetization,Katou19}. In the case of S2, \cite{Jebaraj24} found that the transition consisted of a series of large-amplitude quasi-perpendicular waves, with a wide range of electron-scale waves present in the foreshock. This combination makes it difficult to pinpoint a single mechanism, though adiabatic, non-adiabatic, and other diffusive mechanisms could coexist.

\textit{September 5, 2022 (S1)}: To directly compare the accelerated electrons and the synchrotron radiation they emit, we fit the emission peaks \(f_\mathrm{c}\), marked with distinct indicators in Figure \ref{fig1}A, using power laws. For the same time interval, we present the time evolution of the relativistic electron intensity (in unprocessed units of counts s\(^{-1}\)) measured by EPI-Hi/HET. The relativistic electron intensity rapidly increases approximately 5 minutes before the shock crossing and remains substantial and nearly constant up to the crossing itself. This suggests that relativistic electrons are found in a relatively small region upstream of the shock. In the post-shock region, the relativistic electron intensity decreases significantly, suggesting that emission must be produced by electrons in the upstream. We construct the electron energy spectra shown in Figure \ref{fig2}B using data collected by the EPI-Lo and EPI-Hi/HET instruments during the shock crossing, between 17:17 and 17:27 UT. Both detectors of EPI-Hi/HET, namely, the Sun-facing HET-A and the anti-Sun-facing HET-B, are used. Unlike in Figure \ref{fig2}A, the data used to construct the energy--intensity spectra are processed in units of intensity, i.e., particles cm\(^{-2}\) s\(^{-1}\) sr\(^{-1}\) keV\(^{-1}\). We use a 10-minute average of the data to reduce large uncertainties that may arise due to low statistics. Detailed information about the unfolding procedure used to construct the energy spectra and the associated errors is provided in Appendix \ref{app:ISIS}.

The energy spectra shown in Figure \ref{fig2}B span from sub-relativistic energies, where \(\gamma \sim 1\) or \(E_\mathrm{e} \sim 50\) keV, to 4 MeV (or \(\gamma \sim 8\)). Due to missing energy channels between 200 keV and 500 keV, we approximately fit a single power law from 50 keV up to 1 MeV, yielding an exponent \(\delta \sim 2.85\). However, these electrons, with \(\gamma \sim 1 - 2\), do not contribute significantly to synchrotron power, which scales as \(\mathcal{P} \sim \gamma^2\). Beyond 1 MeV, uncertainties in HET measurements increase (see Appendix \ref{app:ISIS}), making precise determination of \(\delta\) challenging. Despite these uncertainties, we observe a deviation from the power law obtained for lower energies beyond 1 MeV. A new power law fit, combining data from HET-A and HET-B between 1 and 5 MeV, yields an exponent \(\delta \sim 4.5\). However, when assessed separately, HET-A and HET-B do not follow the same trend and can be fit with separate power laws, with exponents \(\delta \sim 4.4\) for HET-A (shown in red) and \(\delta \sim 5\) for HET-B (shown in green). 

Isolating the theoretically predicted shape of the synchrotron spectra is impossible in the presence of elevated background radiation from thermal and nonthermal sources during PSP's close encounters \citep[][]{Liu23}. Therefore, in what follows we focus on identifying \(f_\mathrm{c}\) as seen in Figure \ref{fig1}A and estimating \(\alpha\) in the optically thin regime to infer the corresponding \(\delta\). Figure \ref{fig2}C shows temporal cuts of spectral power as a function of frequency at selected times between 16:42 UT and the shock arrival at 17:27 UT. We focus on Syn-2 (square marker) and Syn-3 (diamond marker), as Syn-1 (circle marker) reaches the local \(f_\mathrm{pe}\) early and is not relevant to the locally measured electrons. For Syn-2 and Syn-3, \(f_\mathrm{c}\) is well-defined by intensity peaks, allowing for the estimation of the spectral slope \(\alpha\) above \(f_\mathrm{c}\). Between 16:42 UT and 17:15 UT, we find \(\alpha \sim 2\) (\(\delta \sim 5\)) for both Syn-2 and Syn-3, consistent with the \(\delta \sim 4.5\) estimated from the \textit{in situ} electron measurements (Figure \ref{fig2}B). After 17:15 UT, as \(f_\mathrm{c}\) approaches \(f_\mathrm{pe}\), the minimum electron energy \(\gamma_\mathrm{min}\) required to emit above \(f_\mathrm{pe}\) increases. In this case, emission originates from the steep tail of the electron distribution resulting in a steep \(\alpha \gtrsim 8\) or \(\delta \gtrsim 17\). Alternatively, this may be a result of the high frequency exponential synchrotron cut off, \(f_\mathrm{max} = \omega_\mathrm{max}/2\pi\) at \(\gamma_\mathrm{max}\) of the electron distribution. 

Interpreting these observations from a DSA perspective, we find that the power law exponent of the low-energy electrons \(\delta \sim 2.85\) aligns reasonably well with the prediction for S1, \(\delta \sim 3\). However, DSA is not particularly effective for the acceleration of sub-relativistic electrons, and it is likely that only electrons with energy \(E_\mathrm{e} \gtrsim 1\) MeV can efficiently participate in the process. The observations presented here show that at 1 MeV and above, the power law exponent steepens to \(\delta \sim 4.5\), suggesting that even if DSA is active, it is inefficient. The steepening of \(\alpha\) also suggests that there is a cutoff in DSA close to \(E_\mathrm{e} \sim 4\)~MeV. Here, we find varying electron intensities measured by HET-A and HET-B above \(\gtrsim 1.5\) MeV. When the magnetic field is nominally oriented, the Sun (HET-A) and anti-Sun (HET-B) directions correspond to parallel (\(0^{\circ}\)) and anti-parallel (\(180^{\circ}\)) pitch angles of the electrons. Since the electron intensity measured by HET-A is more than an order of magnitude higher than HET-B, it indicates strong anisotropy in the electrons observed from the Sun direction, where the shock is approaching the observer. Anisotropy is generally found when the accelerated particles do not partake in efficient DSA or in the absence of a well-developed foreshock. 

\begin{figure*}[ht!]
\centering
\includegraphics[width=0.65\textwidth]{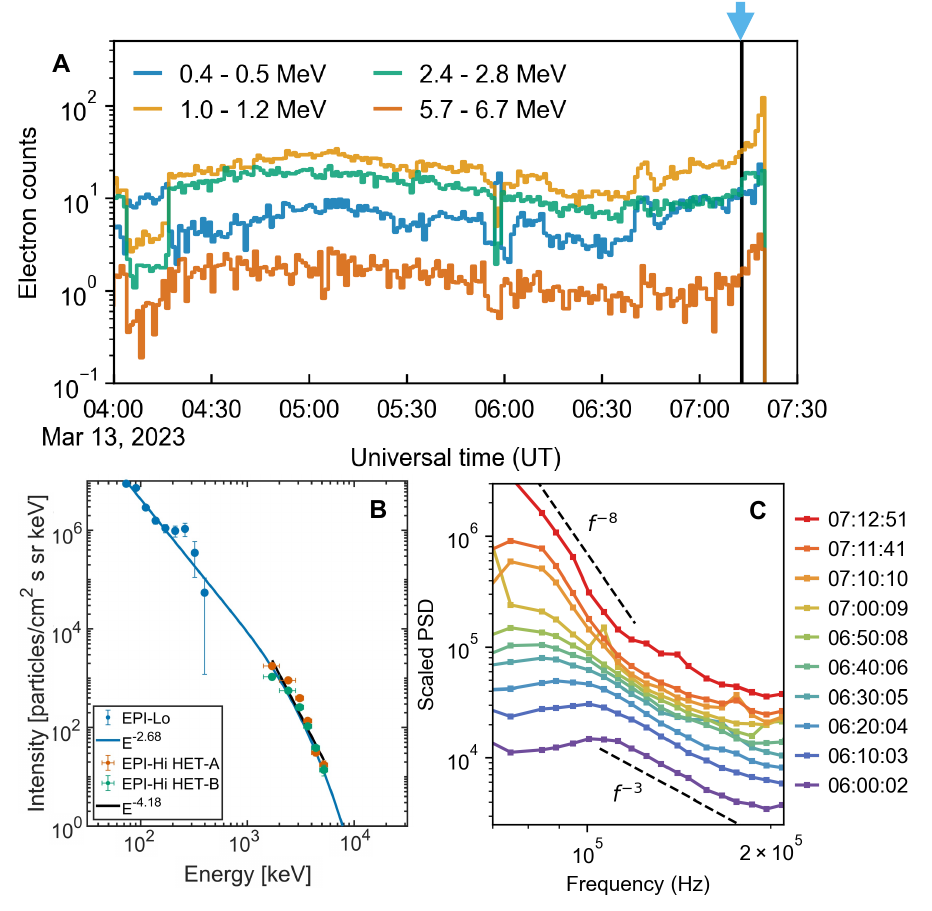}
\caption{Characteristics of the synchrotron emission from S2 and the corresponding electron distribution. Panel A presents the time evolution of the electron intensity in units of counts s\(^{-1}\) observed between 04:00~UT and 07:30~UT. The arrival of the shock at PSP is marked by the vertical black line at 07:13~UT and the blue arrow on top. Panel B shows the 10-minute averaged electron energy distribution between 07:03-07:13~UT in units of particles cm\(^{-2}\) s\(^{-1}\) sr\(^{-1}\) keV\(^{-1}\). The different detectors of IS\(\odot\)IS used to construct the electron spectrum are annotated with colored markers. The approximate power law fits and the exponents are also shown in the legend. Panel C shows the intensity-frequency profiles at distinct times leading up to the shock arrival, with approximate power law fits to the optically thin regime is shown.}\label{fig2_2}
\end{figure*}

\textit{March 13, 2023 (S2)}:We present the temporal evolution of the relativistic electron population in Figure \ref{fig2_2}A, which shows an increase in particle counts leading up to the shock arrival. Unlike in S1, the relativistic electron intensity in S2 is relatively high, with a noticeable increase starting \(\sim\)06:40 UT and continuing until the shock's arrival at 07:13 UT. Figure \ref{fig2_2}B illustrates the average electron energy--intensity spectrum in a 10-minute interval upstream of the shock arrival. Using a similar power law fitting as for S1, we found an exponent of \(\delta \sim 2.7\) between 50 keV and 2.5 MeV (\(\gamma \sim 4\)). Unlike S1, the power law extends beyond sub-relativistic energies up to \(\gamma \sim 4\), making it more relevant for synchrotron emission. Beyond \(\sim\)2.5 MeV, the spectrum transitions into a much steeper power law or exhibits an exponential rollover. Despite increased uncertainties due to the instrument's dynamic threshold mode (see Appendix \ref{app:ISIS}), we approximate a power law fit for this data and obtain an exponent of \(\delta \sim 4.18\).

Next, we isolate the maxima \(f_\mathrm{c}\) of the synchrotron emission, shown with circle markers in Figure \ref{fig1}B, and construct the spectra of the optically thin emission above \(f_\mathrm{c}\), as displayed in Figure \ref{fig2_2}C. We follow the evolution of this spectrum over a 72-minute period from 06:00 to 07:12 UT, up until the shock arrives at the observer. Between 06:00 UT and 07:00 UT, \(\alpha\) remains constant at approximately 3 (\(\delta \sim 7\)), which is steeper than the \(\delta \sim 4.18\) measured \textit{in situ}. This discrepancy may arise from large uncertainties in the \(\delta\) estimated from the data in Figure \ref{fig2_2}B. Alternatively, it could indicate that the emission is produced by electrons in the steeper part of the electron distribution, such that \(f_\mathrm{c}\) corresponds to \(\gamma_\mathrm{min} > 4\) or \(E_\mathrm{e} \sim 2\) MeV, where the power law breaks in Figure \ref{fig2_2}B. After 07:00 UT, as the shock approaches the observer and \(f_\mathrm{c}\) approaches \(f_\mathrm{pe}\), \(\gamma_\mathrm{min}\) increases, further steepening \(\alpha\) to approximately 5 (\(\delta \sim 11\)). Between 07:10 UT and 07:12 UT, just before the shock arrives and when the observer is within the emitting region, \(\alpha \gtrsim 8\) (\(\delta \gtrsim 17\)). Similar to S1, this may be a result of the high frequency exponential synchrotron cut off at \(\gamma_\mathrm{max}\).

Unlike S1, we can directly compare DSA predictions with observations, as the power law extends up to \(\gamma \sim 5\), the energy range where electrons are affected by the DSA process. DSA predicts \(\delta = 2\), which differs from the observed power law exponent of \(\delta \sim 2.7\), suggesting the presence of additional factors that reduce DSA efficiency. Regarding anisotropy, both HET-A and HET-B detect similar intensities and energies up to \(\sim 6.7\) MeV (\(\gamma_\mathrm{max} \sim 13\)). This indicates that the electrons are isotropic, likely due to the presence of a well-developed foreshock that enables efficient DSA. A final important observation from Figure \ref{fig2_2}B is the order-of-magnitude higher intensity of the entire population of electrons, from sub-relativistic to relativistic, compared to what was estimated for S1 in Figure \ref{fig2}B. 

\section{Discussion}

In this study, we present the first-ever \textit{in situ} observations of shocks that produced an accelerated power-law distribution of electrons with \(\gamma \gg 1\), capable of emitting synchrotron radiation. Unlike the previous remote-sensing study by \cite{Bastian07}, we experimentally verified the properties of synchrotron emission and the conditions under which it is generated. This was made possible by the unique trajectory of the PSP, which, for the first time, allowed us to study the strongest and fastest IP shocks near their origin at the Sun. Our detailed analysis of both the relativistic electron distribution and the characteristics of the synchrotron radiation led to several key findings, summarized below.

We first demonstrated that the observed radiation characteristics, including spectral morphology and brightness temperature, align with theoretical predictions. For S1, we measured circular polarization in the x-mode, consistent with plasma theory, and found further agreement in the relationship between circular polarization and the \(\gamma\) of the emitting electrons. However, both S1 and S2 exhibited significant deviations from the expected maximum polarization. The theoretical maximum polarization, \(\Pi = \frac{3\delta + 3}{3\delta + 7}\), using \(\delta\) values from S1 and S2, yields \(\Pi \sim 80\%\). Notably, emission from S2 was completely depolarized, suggesting that in strongly coupled plasma with multi-scale inhomogeneities (both parallel and perpendicular to \(\mathbf{B}\)), such as those found in the solar wind or interstellar medium, the emitted electromagnetic wave characteristics may be lost near the emission region. This result is consistent with recent X-ray polarimetric studies of SNRs \citep{Vink22}. These studies show only a small degree of polarization due to a mixture of tangential magnetic-field alignment near the shock front and a radially aligned magnetic-field structure further from it \citep{Bykov20}.

Secondly, both shocks exhibited similar power-law exponents (\(\delta \sim 2.7-2.9\)) in the low-energy regime before breaking or transitioning to an exponential cutoff. S1 showed partial agreement with DSA predictions for \(r_\mathrm{gas} \sim 2.5\), yielding \(\delta \sim 3\). However, because electrons with \(\gamma \gtrsim 2\) are required to participate in DSA, and the power-law did not extend beyond \(\gamma \sim 2\) for S1, it is unclear if DSA was fully active or efficient. Additionally, we observed significant anisotropy, likely due to minimal pitch angle scattering -- which suggests that DSA was inefficient. For S2, we found \(\delta \sim 2.7\), steeper than the DSA prediction of \(\delta = 2\) for \(r_\mathrm{gas} \sim 4\). Despite this, the power-law extended up to \(\gamma \sim 5\) before breaking. The electron distribution in S2 was isotropic, a hallmark of efficient DSA. 

Lastly, our analysis of the \textit{in situ} relativistic electron distributions revealed that a quasi-parallel shock like S2 is far more efficient at accelerating electrons than a quasi-perpendicular shock like S1. Electron intensities near the shock where synchrotron emission was generated were an order of magnitude higher for S2 than for S1, resulting in much stronger synchrotron radiation from S2. A plausible explanation is that in quasi-perpendicular shocks, adiabatic acceleration mechanisms likely dominate, resulting in low efficiency for producing relativistic particles \citep[e.g.,][]{Jebaraj23l}. This suggests that additional mechanisms are required for producing larger populations of relativistic electrons. If DSA were the additional mechanism, its efficiency would be limited by the growth of ion-scale waves through ion-streaming instabilities, a process that is highly inefficient in quasi-perpendicular shocks like S1 \citep{Zank92}. In contrast, \cite{Jebaraj24} found that S2 can locally provide quasi-perpendicular conditions while supporting the efficient growth of ion-scale waves due to its large-scale quasi-parallel geometry making it an efficient accelerator of relativistic electrons.

The last result also experimentally confirms what was observed in the bilateral supernova remnant SN 1006, where the analysis of remote-sensing observations demonstrates that synchrotron emissions from quasi-parallel SNR shocks were significantly brighter than those from oblique and quasi-perpendicular shocks (e.g., \citealt{Rothenflug2004}, \citealt{Giuffrida22}). This consistency is expected, as the structure of quasi-parallel shocks align with theoretical predictions \citep{Balikhin23}. Consequently, the non-thermal radiation resulting from the evolutionary behavior of quasi-parallel SNR and IP shocks is likely similar.

The experimental results presented here are the first of their kind, showing that certain IP shocks are capable of persistently accelerating electrons to relativistic energies—a phenomenon previously considered unlikely. The mechanisms by which these shocks accelerate particles and produce radiation resemble those seen in SNR shocks. While detailed quantification of differences between IP and SNR shocks is beyond the scope of this letter, our work suggests that efficient acceleration is directly related to the upstream bulk flow energy dissipated at the shock in its rest frame.

\subsection*{Acknowledgments} 
\textbf{Author contributions:} The study was initiated by ICJ and conceptualized together with OVA, VVK, MG, and MM. The methodologies for data analysis was provided by ICJ, OVA, LV, MG, AV, and RV. Data analysis was performed by ICJ, OVA, LV, and AV. The results were interpreted by ICJ, OVA, MG, RV, MM, VVK, JP, and MB. The data was visualized by ICJ, OVA, LV, and AV. AM prepared the schematic. ICJ wrote the manuscript and prepared the final draft with significant revisions from MG. All authors have read the manuscript and agreed to the presented results. The science operations of PSP are led by NER. The operations of the IS\(\odot\)IS instrument suite is led by DJM, and the data was processed by CMSC, JGM, and AL. The operation of FIELDS instrument suite is led by SDB and MP.

\textbf{Funding information:}
The Parker Solar Probe spacecraft was designed, built, and is now operated by the Johns Hopkins Applied Physics Laboratory as part of NASA’s Living with a Star (LWS) program (contract NNN06AA01C). Support from the LWS management and technical team has played a critical role in the success of the Parker Solar Probe mission.
The authors express their gratitude to all the instrument teams for their work in processing and publishing the publicly available data from the Parker Solar Probe. The data used in this study are available at the NASA Space Physics Data Facility (SPDF), https://spdf.gsfc.nasa.gov.
This research was supported by the International Space Science Institute (ISSI) in Bern through ISSI International Team project No.~23-575, ``\textit{Collisionless Shock as a Self-Regulatory System}''. This research was supported through the Visiting Scientist program of the International Space Science Institute (ISSI) in Bern.
ICJ, LV, and ND are grateful for support by the Research Council of Finland (SHOCKSEE, grant No.~346902), and the European Union’s (E.U's) Horizon 2020 research and innovation program under grant agreement No.~101004159 (SERPENTINE) and No.\ 101134999 (SOLER). The study reflects only the authors' view and the European Commission is not responsible for any use that may be made of the information it contains.
OVA was partially supported by NSF grant number 1914670, NASA’s Living with a Star (LWS) program (contract 80NSSC20K0218), and NASA grants contracts 80NNSC19K0848, 80NSSC22K0433, 80NSSC22K0522. OVA and VVK were supported by NASA grants 80NSSC20K0697 and 80NSSC21K1770.
LV acknowledges the financial support of the University of Turku Graduate School. 
VVK also acknowledges financial support from CNES through grants ``Parker Solar Probe'' and ``Solar Orbiter''.
AK acknowledges financial support from NASA NNN06AA01C (PSP EPI-Lo) contract.
NW acknowledges funding from the Research Foundation -- Flanders (FWO -- Vlaanderen, fellowship no.\ 1184319N).
EP acknowledges support from NASA's Parker Solar Probe Guest Investigators (PSP-GI; grant no.~80NSSC22K0349) and Living With a Star (LWS; grant no.~80NSSC19K0067) programs.
JP acknowledges support from the Research Council of Finland (SWATCH, grant No. ~343581).
%
%
%
The FIELDS experiment was developed and is operated under NASA contract NNN06AA01C.
\\


\appendix


\section{Experimental details} \label{App:sec1}

\subsection{FIELDS}

Our study primarily utilizes the high-frequency electric field measurements from both the high-frequency and low-frequency receivers (HFR \& LFR) of the Radio Frequency Spectrometer \citep[RFS;][]{Pulupa2017}. These measurements are made by the FIELDS instrument suite on board the PSP spacecraft \citep{Bale16}. RFS includes four electric antennas (\(\vec{V}_1, \vec{V}_2, \vec{V}_3, \text{and}~\vec{V}_4\)) and measures over a wide frequency range, spanning from 20 MHz to 1 kHz at a 3.5 second cadence during close encounters. The frequency resolutions is 4\% at any given time over 64 channels in LFR and 64 in HFR.

Additionally, PSP measures the full EM fields for which, the coordinate system used throughout the manuscript is the inertial RTN (Radial--Tangential--Normal) system, where the radial component R is oriented along the Sun--spacecraft line, the transverse component T is defined to be orthogonal to the rotational axis of the Sun and the radial component, i.e., \(T = \Omega_{\odot} \times R\), while the normal component N completes the orthogonal right-handed triad and, in this case, is aligned with the normal of the ecliptic plane. The electric field measurements are made using the electric fields instrument (EFI) consisting of two pairs of dipole electric field antennas oriented in the TN plane and extending beyond the PSP heat shield, and a fifth antenna located behind the heat shield on the instrument boom; the location of antenna \(\vec{V}_5\) in the wake of PSP means the R component is susceptible to detrimental interference by the wake electric field and cannot be reliably interpreted \citep{Bale16}. Two three-component flux-gate magnetometers (MAG) measure the magnetic field from DC up to 293 vector measurements per second during 2 to 4 days around close encounter. The latter is used in the present study. 

The Time Domain Sampler (TDS) provides high-frequency measurements of waveforms at a rate of about 1.92 million samples s\(^{-1}\). Waveform events captured by TDS consist of ~15 milliseconds of the electric (2 components) and magnetic (1 component) field measurements \citep{Bale16}. TDS recorded 66 events on September 5, 2022 between 16:00 and 17:30~UT, and 66 events on March 13, 2023 between 04:00 and 07:30~UT. These events were used to compliment the HFR/LFR data. 

\subsection{IS\(\odot\)IS} \label{app:ISIS}

To analyze the energetic electron spectra close to the shock, we used the Integrated Science Investigation of the Sun, IS\(\odot\)IS, instrument suite \citep{McComas16}. It measures energetic particles from $\sim$20 keV to over 100~MeV/nuc with two Energetic Particle Instruments (EPI), EPI-Lo \citep{Hill17} and EPI-Hi \citep{Wiedenbeck17}. EPI-Lo measures energetic electrons primarily through the ``ChanE'' data product which utilizes a single silicon solid-state detector (SSD) for the energy measurement and a secondary SSD in anticoincidence with the front SSD to reject penetrating electrons.  The High Energy Telescope (HET) within EPI-Hi provides measurements of higher energy electrons with minimal contamination from incident ions.  HET is a double-ended dE/dx vs. residual energy telescope with one aperture (HET-A) pointed toward the Sun along a nominmal Parker spiral and HET-B pointing the opposite direction.  For both EPI-Lo and EPI-Hi/HET, the measured electron spectra were unfolded using the response matrix technique to account for the contribution of higher energy electrons being measured in a lower energy bin.  Full details of the Monte Carlo simulations utilized to construct these response matrices are provided in \cite{2022PhDT........23M} and \cite{Labrador:2023lv}.  While EPI-Hi/HET electron measurements are essentially unaffected by ion contamination, EPI-Lo ChanE can have substantial contamination from high energy ions when there is a significant SEP ion foreground \citep{2021ApJ...919..119M}.  The EPI-Lo response matrix factors in contamination from high energy ions and reliably separates them from theelectron signals \citep{2022PhDT........23M}.  As such, only the filled markers in Figure \ref{fig2} should be interpreted as being produced by electrons.

During normal operation, all detector segments in EPI-Hi/HET are sensitive to protons, alphas, electrons, and heavier ions.  In periods of high energetic particle intensity, the EPI-Hi instruments employ automated ``dynamic threshold'' states which raise the trigger thresholds of some detector segments such that they are sensitive only to $Z\geqq6$ ions.  This lowers the effective geometry factor of the instrument for protons, alphas, and electrons, and keeps the instrument livetime from dropping to low levels, while also preserving the ability to measure heavy ions during large events.  In the case of the most restrictive dynamic threshold state, HET electron sensitivity is reduced in energy range to $\sim2$ to $\sim5$ MeV.

\section{Transverse electromagnetic wave properties} \label{app:TDS}

\begin{figure}[h]
\centering
\includegraphics[width=0.5\textwidth]{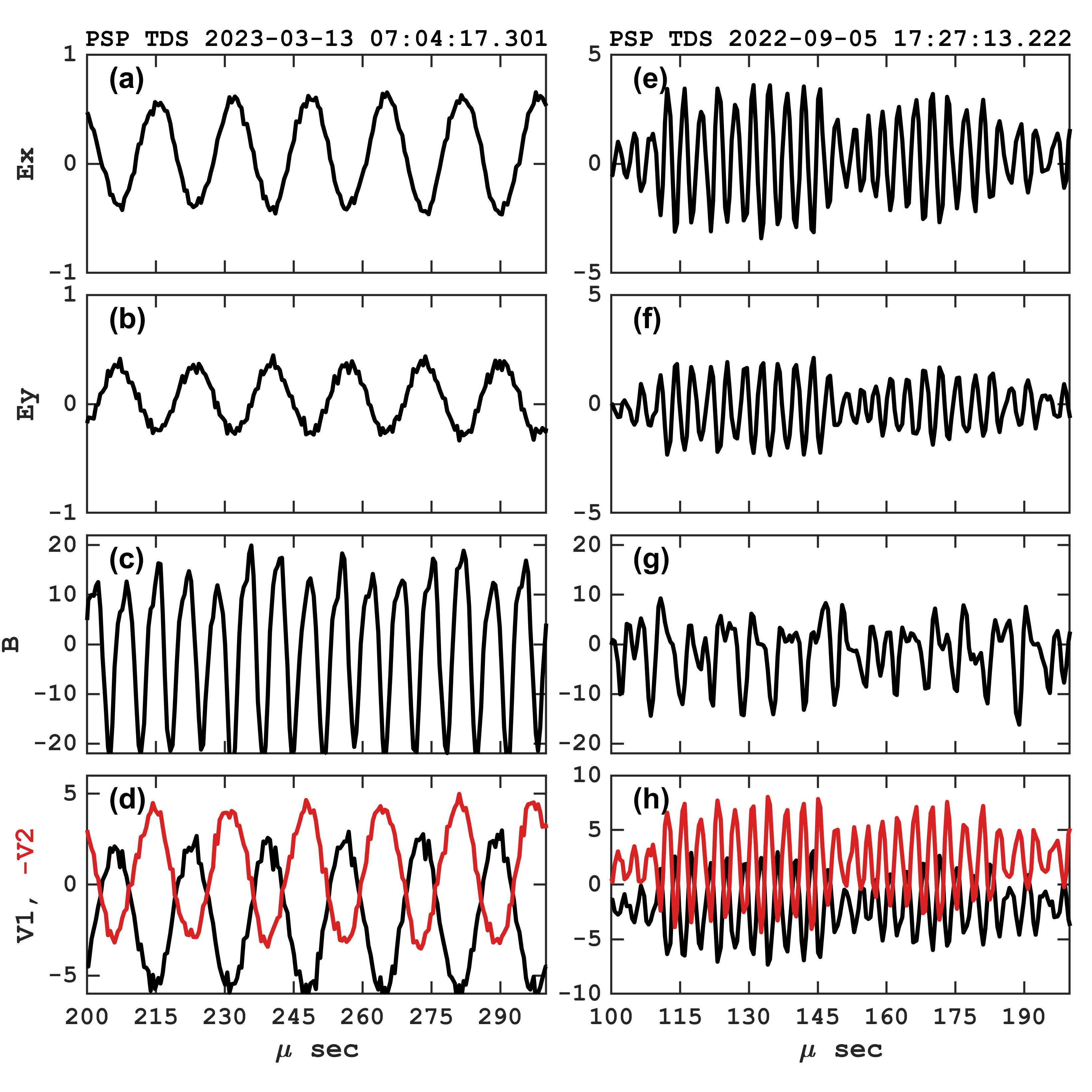}
\caption{The potential waveforms recorded by TDS on March 13, 2023, 07:04:17.301: (a),(b) - two components of the electric field; (c) - the magnetic field; (d) the potential recorded on antenna 1 (black) and 2 (red). Panels (e)-(h): waveforms recorded by FIELDS/TDS on September 5, 2022, 17:27:13.222.
}
\label{TDS}
\end{figure}

Figure \ref{TDS} displays a waveform captured by the FIELDS/TDS on March 13, 2023. The electric field measurements depicted in panels (a) and (b) reveal the presence of a wave with a frequency of approximately 60 kHz, which aligns with the synchrotron emission illustrated in Figure \ref{fig1}. Magnetic field measurements in panel (c) exhibit only 150 kHz noise, indicating that the magnetic aspect of the wave is too faint for detection by the Search Coil Magnetometer (SCM).

For further insights into the wave characteristics, we have performed processing of the plasma potential recorded by \(\vec{V}_\mathrm{1}\) and \(\vec{V}_\mathrm{2}\) in panel (d). Notably, the absence of a phase difference in the time profiles allows for estimations of the lower limits of the wavelength and phase velocity of the recorded wave. Given the separation distance between the antennas, denoted as $L\sim$ 4 m, and the time step $\Delta t\sim$ 0.5$\times$10$^{-6}$ seconds, the wavelength is estimated to exceed $L/(f \Delta t)\sim$ 133 m. Consequently, the resulting phase velocity greatly exceeds $\sim$ 7.8$\times$ 10$^{3}$ km s\(^{-1}\). These derived values surpass those anticipated for any electrostatic plasma wave under the prevailing plasma conditions. Analysis of the waveforms recorder on September 5, 2022, gave similar results. 
Example of wave with a frequency of approximately 420 kHz (consistent with the synchrotron emission illustrated Figure \ref{fig1}) is shown in panels (e)-(h).

\section{Estimating Brightness Temperature} \label{app:Brightnesstemp}

The brightness temperatures of synchrotron sources have an upper limit at low frequencies due to synchrotron self-absorption, limiting the maximum brightness temperature to that of the kinetic temperature of the emitting electrons. For a nonthermal source where the electron energy distribution follows a power law, the brightness temperature cannot exceed the effective temperature of the relativistic electrons responsible for the emission at a given frequency. 

At sufficiently low frequencies, the brightness temperature \(T_\mathrm{B}\) approaches the effective electron temperature \(T_e\), calculated as:
\begin{equation}
    T_e \sim \left( \frac{2\pi m_\mathrm{e} c f}{eB} \right)^{1/2} \frac{2 m_\mathrm{e} c^2}{3k_\mathrm{B}}.
\end{equation}
The brightness temperature in the Rayleigh-Jeans limit is defined as:
\begin{equation}
    T_\mathrm{B} = \frac{I_{f}c^2}{2k_\mathrm{B} f^2}.
\end{equation}
For an optically thick synchrotron source, \(T_\mathrm{B}\) cannot exceed \(T_e\), and the flux density \(S\) relates to \(T_\mathrm{B}\) through:
\begin{equation}
   S = 2k_\mathrm{B} T_\mathrm{B} f^2 \Omega / c^2,\label{eq:brithnesstempsolid}
\end{equation}
where \(\Omega\) is the solid angle subtended by the source. In order to estimate the brightness temperature of the radiation, this solid angle needs to be estimated. The solid angle for distant astrophysical sources is approximated to be \(\Omega = \pi \theta^2\), but for sources much closer, this approximation falls apart. For large angles subtended by the source, the \(\Omega = 2 \pi (1-\cos{\theta})\) \citep[][]{Chandrasekar60}. Here, we estimate $\theta = \arctan{(s/D)}$ by taking the distance between the shock at $r\sim V_\mathrm{shock}\Delta t$ and the observer at $r_\mathrm{sc}$ as $D = r_\mathrm{sc}-r$ and the source width as $s\sim r$. For S1 we assume $V_\mathrm{shock}=1800\,$km/s and for S2, $V_\mathrm{shock}=2500\,$km/s based on simple transit time estimations. 

\begin{figure}[ht!]
\centering
\includegraphics[width=0.5\textwidth]{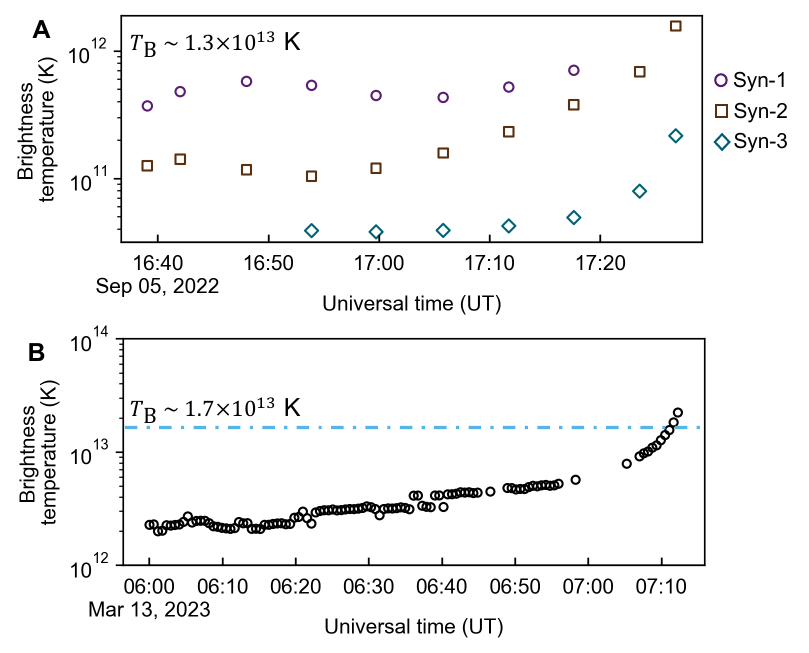}
\caption{The brightness temperature (\(T_\mathrm{B}\)) of the synchrotron emission for S1 and S2 are shown in panel A, and B. The inverse-Compton scattering limit is explicitly stated in both panels and also represented by the blue dot-dashed in panel B. }\label{figA2}
\end{figure}

\(T_\mathrm{B}\) of a synchrotron source that absorbs its own radiation is restricted by inverse Compton scattering, with a theoretical maximum around \(10^{12}\) Kelvin, as noted by \cite{Kellerman69}. This limit is given by the ratio of inverse Compton to synchrotron cooling depends primarily on \( (d f T_\mathrm{B}^5) \), being highly sensitive to brightness temperature. Here 
\begin{equation}
    d = \frac{1}{1 - \alpha} \left(\frac{f_{\mathrm{max}}}{f_\mathrm{c}}\right)^{1-\alpha},
\end{equation}
where \( \alpha \) is the spectral index of the synchrotron spectrum, \( f_{\text{max}} \) is the maximum frequency of the synchrotron emission, and \( f_\mathrm{c} \) is the turnover frequency below which self-absorption is active. \(d\) is a dimensionless parameter which is of order 10 for astronomical sources and becomes critical for determining the inverse Compton scattering limit. For typical values of \( d \) and \( f \), the brightness temperature limit to prevent the Compton catastrophe is:
\begin{equation}
T_\mathrm{B} < 1.5 \times 10^{12} \left( \frac{f}{10^9~\text{Hz}} \right)^{-1/5} \left( \frac{d}{10} \right)^{-1/5} \text{K}.
\end{equation}

It should be noted that the \(10^{12}\) K limit is given for 1 GHz and for \(d\) of order 10, and both \(f\) and \(d\) scale weakly as \(f^{-1/5}\), and \(d^{-1/5}\). Given that the frequency at which the synchrotron radiation is observed is at \(\sim 4\) orders of magnitude smaller, the limit becomes \(\approx 10^{13}\) K. 

Assessing the \(T_\mathrm{B}\) for Syn-1, Syn-2, and Syn-3 in Figure \ref{figA2}A, we find that Syn-1 maintains a nearly constant \(T_\mathrm{B}\), while the \(T_\mathrm{B}\) of Syn-2 and Syn-3 increase rapidly prior to shock arrival at 17:27 UT. However, this increase does not exceed the inverse Compton scattering limit of \(1.3 \times 10^{13}\) K, with \(f = 400\) kHz, and for \(\alpha = 3\). Figure \ref{figA2}B shows the brightness temperature \(T_\mathrm{B}\) of the emission, which rises exponentially as the shock nears the spacecraft. The \(T_\mathrm{B}\) even exceeds the inverse Compton limit of \(1.7\times10^{13}\) K at 07:10, suggesting that \(\Omega = 2\pi\), at which point the estimation of \(T_\mathrm{B}\) becomes unreliable. This indicates, that at 07:10~UT the spacecraft should have been in the region where the electrons emitting synchrotron radiation were being accelerated.

The limit is widely accepted and is important to distinguish synchrotron radiation which in incoherent from coherent emissions. The defining characteristic of coherent emissions is their very high intensities \citep[][]{Suzuki85book,Krasnoselskikh19}, which in \(T_\mathrm{B}\) units greatly exceeds \(10^{12}\) K and can reach \(10^{18}\) K \citep[][]{saint2012decade}. This suggests that coherent emissions have far greater \(T_\mathrm{B}\) than what is allowed for a hypothetical black body radiation of the emitting electrons. Estimating and analyzing the \(T_\mathrm{B}\) can help in the identification of the emission mechanism. When \(T_\mathrm{B}\) exceeds the limits found here, it rules out synchrotron radiation as their primary emission process. 

\section{Polarization} \label{app:stokes}
A monochromatic electromagnetic wave of the frequency \(\omega\) is observed, at a receiver, by measuring the electric field \(\bm{E}\) which lies in the plane, perpendicular to the line of sight, i.e., the direction of propagation. This electric field is conveniently represented as 
\(\bm{E}=\real (E\hat{\bm{e}} e^{-i\omega t}).\) The real amplitude \(E\) determines the intensity, while the complex unit polarization vector \(\hat{\bm{e}}\) fully determines the axes of the polarization ellipse and the direction of the electric vector rotation for the completely polarized wave. If the emission comes from many sources, the receiver measures a superposition of many waves 
\begin{align}
&\bm{E}=\sum_i \real(E_i\hat{\bm{e}}_ie^{-i\omega t})
\end{align}
where the summation is over the ensemble of the emitters. Since all \(E_i\) and \(\hat{\bm{e}}_i\) may be different and independent, it is not possible to describe the emission with a single polarizarion axis. Instead, the Stokes parameters are widely used~\citep{chandrasekhar1947transfer, Rybicki79}. These parameters are the differences of average intensities projected onto several sets of unit vectors. Let the plane of the electric field be the \(x-y\) plane. Consider the following pairs of the unit vectors: 
a) \(\hat{\bm{e}}_x=\hat{x}\), \(\hat{\bm{e}}_y=\hat{y}\); b) \(\hat{\bm{e}}_a=(\hat{x}+\hat{y})/\sqrt{2}\), \(\hat{\bm{e}}_b=(\hat{x}-\hat{y})/\sqrt{2}\); and c) 
\(\hat{\bm{e}}_R=(\hat{x}+i\hat{y})/\sqrt{2}\),\( \hat{\bm{e}}_L=(\hat{x}-i\hat{y})/\sqrt{2}\). Then the Stokes parameters are defined as follows:
\begin{align}
&I=\expval{|\bm{E}\cdot\hat{\bm{e}}_x|^2} + \expval{|\bm{E}\cdot\hat{\bm{e}}_y|^2}\\
&Q=\expval{|\bm{E}\cdot\hat{\bm{e}}_x|^2} - \expval{|\bm{E}\cdot\hat{\bm{e}}_y|^2}\\
&U=\expval{|\bm{E}\cdot\hat{\bm{e}}_a|^2} - \expval{|\bm{E}\cdot\hat{\bm{e}}_b|^2}\\
&V=\expval{|\bm{E}\cdot\hat{\bm{e}}_R|^2} - \expval{|\bm{E}\cdot\hat{\bm{e}}_L|^2}
\end{align}
where \(\expval{\ldots}\) means ensemble averaging which corresponds to the averaging of the observations over the time much larger than \(1/\omega\). Accordingly, the linear polarization degree is \(\Pi_\mathrm{lin}=\sqrt{Q^2+U^2}/I\), the circular polarization degree is \(\Pi_\mathrm{cir}=V/I\), and the total polarization degree is \(\Pi_\mathrm{tot}=\sqrt{Q^2+U^2+V^2}/I\).

This final expression, \(\Pi_\mathrm{tot}\), is important since it is used in observational astronomy to determine the characteristics of the cosmic object and the medium in which it is situated \citep[][]{Bykov12}. In radio astronomy, Stokes parameters are often calculated by converting incident electric fields into voltages, where the detection process yields power proportional to the square of these voltages. The raw data, initially in arbitrary units, is calibrated to brightness temperature (K) or flux density (W m\(^{-2}\) Hz\(^{-1}\)), with a constant factor accounting for intensity calibration. The exact details of how PSP data is calibrated and the polarization is estimated can be found in \cite{Pulupa2017}. 

\bibliography{bibTeX_11_11_2022}{}
\bibliographystyle{aasjournal}

\end{document}